\documentclass[aps,amsmath,pra,reprint]{revtex4-1} 
\usepackage{epsf}
\usepackage{graphicx}  
\usepackage{amsmath} 
\usepackage{amssymb} 
   
\newcommand{\be}{\begin{equation}} 
\newcommand{\ee}{\end{equation}}
\newcommand{\eps}{\epsilon}

\begin{document}

\title{A multilayered effective medium model for the roughness effect on the Casimir force}

\author{Andr\'e Gusso} 
\email{gusso@metal.eeimvr.uff.br/andre.gusso@pq.cnpq.br}
\author{\'Ursula Berion Reis}
\affiliation{Departamento de Ci\^encias Exatas-EEIMVR,
Universidade Federal Fluminense, Volta Redonda, 27255-125, RJ, Brazil.}

\begin{abstract}
A multilayered effective medium model is proposed to calculate the contribution of surface roughness to the Casimir force. In this model  the rough layer has its optical properties derived from an effective medium approximation, with the rough layer  considered as the mixing of voids and solid material. The rough layer can be divided into sublayers consisting of different volume fractions of voids and solid material as a function of the roughness surface profile. The Casimir force is then calculated using the generalizations of the Lifshitz theory  for multilayered planar systems. Predictions of the Casimir force based on the proposed model are compared with those of well known methods of calculation, usually restricted to be used with large scale roughness. It is concluded that the effect of short scale roughness as predicted by this model is considerably larger than what could be expected from the extrapolation of the results obtained by the other methods.
\end{abstract} 

\maketitle

The calculation of the Casimir force between bodies with microscopically rough surfaces  has been gaining increasing attention in the last few years due to the high precision experiments that have been carried out~\cite{Klimchitskaya09}. So far, only approximate methods of calculation were proposed, which have different and limited ranges of applicability. Two are the most commonly used approaches, the pairwise summation (PWS) and the proximity  force approximation (PFA).  Both approaches are assumed to deliver adequate predictions for  surfaces having large scale roughness, characterized by low spatial frequency (large correlation length $\Lambda$), and small surface gradient. However, the range of applicability of the PFA was shown to extend down to smaller spatial frequencies when compared to PWS~\cite{Bordag09}. More sophisticated  approximated methods, based on more general physical concepts,  have been developed which  have a broader and more clearly defined range of applicability. For instance, we have the methods developed by Maradudin and Mazur~\cite{Maradudin80}, and more recently by Maia Neto {\it et al.}~\cite{MaiaNeto05}. The method proposed by Maia Neto {\it et al.} is based on the scattering approach using a second-order perturbation theory in the roughness amplitude, and it can be applied to surfaces with both the average surface separation $d$ and $\Lambda$ much larger than the rms roughness amplitude $\sigma$. This is qualitatively the same range of applicability of the PFA, however, because the perturbative scattering approach (PSA) takes into account physical contributions not taken into account by the PFA it delivers more reliable results for rough surfaces with smaller spatial frequencies. More precisely, the following inequalities determine the range of applicability of the PWS, $\sigma \ll d \ll \Lambda$,  of PFA, $\sigma, d \ll \Lambda$, and for the PSA it holds that $\sigma \ll \Lambda, d$. These inequalities imply that all these methods can only give reliable results when the modulus of the average surface gradient is much smaller than unity, the restriction being less stringent for the PSA~\cite{ MaiaNeto05}.

In most of the experiments performed so far for the precise measurement of the Casimir force, the microscopic roughness profiles at the surface of the interacting bodies were such that the use of the PFA or PWS was justified~\cite{Bordag09}. However, in a set of recent experiments~\cite{Zwol10} the Casimir force between surfaces with large amplitude short scale roughness was measured, evidencing a large contribution to the force at small separations that could not be accounted for the known methods of calculation. Further motivation for the development of a reliable method to calculate the Casimir force between surfaces having  short scale roughness, characterized by small $\Lambda$, comes from the potential relevance of this force in micro- and nanodevices operating with small gaps~\cite{Bordag09,Gusso,Gusso07,Pruvost09} ranging from tens up to a few hundreds of nanometers. The short scale roughness at the surfaces of such devices results from the limitations in the fabrication processes and are usually characterized by parameters in the nanometer range~\cite{Sundararajan01}. Therefore,  an alternative calculational method is required.

In this work we propose a multilayered effective medium model (MEMM) that is intended to deliver reliable predictions of the Casimir force between surfaces having short scale roughness requiring only a reasonable calculational effort. 

\section{Effective medium model}

The MEMM is based upon two approximations. The first approximation is to consider the rough layer as  an effective medium whose optical properties result from the mixing of voids (vacuum or air) and the material comprising the solid. This effective medium approximation (EMA) has been used successfully over decades to model the optical properties of  rough layers~\cite{Aspnes79,Palik85}. The effective complex dielectric function $\epsilon^{eff}(\omega)$  can, in principle, be calculated using one out of various mixing rules presented in the literature~\cite{Sihvola00,Sihvola08}. We choose to use the Bruggeman mixing rule~\cite{Bruggeman35} which is derived allowing inclusions in a host material of dielectric spheres with random spatial distribution and radius. This choice was motivated by the fact that this mixing rule correctly predicts $\epsilon^{eff}(\omega)$ for any relative fraction of the mixed materials, including either sparse or aggregate random configurations that can mimic the structures actually found at rough surfaces. In addition,  the Bruggeman mixing rule has been shown to be in agreement with the experimental optical data extracted from light reflected from  rough surfaces over a wide range of probed wavelengths~\cite{Aspnes79,Petrik98,Palik85}, the agreement being improved by the use of multilayer models.

For the two-phase mixture considered to model the rough layer the effective complex dielectric function  $\eps^{eff}$ is obtained from the Bruggeman mixing rule
\be
(1-f)\frac{\eps_v-\eps^{eff}}{\eps_v+2 \eps^{eff}}+ f \frac{\eps_s-\eps^{eff}}{\eps_s+2 \eps^{eff}}=0 \,,
\label{Bruggeman}
\ee
where the input parameters $\eps_v$ and $\eps_s$ correspond to the void and solid complex dielectric functions, and $f$ denotes the volume fraction of the solid. For the voids we obviously have $\eps_v=1$. The optical properties of the solid at the rough layer can be approximated by those of the bulk, as has been usually done in the calculation based on other approaches~\cite{Bordag09}, or some other experimental or theoretical dielectric function that is believed to better represent the optical properties of the solid at the rough layer.

\begin{figure}
\centerline{\includegraphics[scale=0.65]{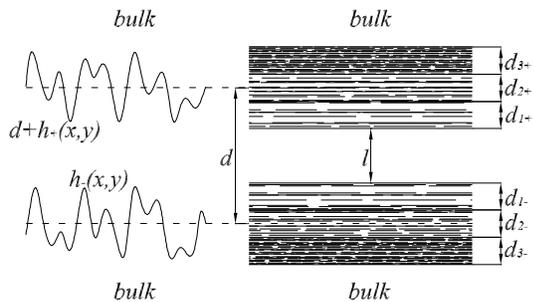}}
\caption{Schematic drawing showing the rough surfaces and an example of a possible effective multilayer model.} 
\label{SchemeLayers}
\end{figure}

For a rough surface characterized by the stochastic function $h(x,y)$, which denotes the deviation in the $z$ direction from the mean value $z=0$, the volume fraction $f$ is generally a function of $z$. The functional dependence of $f$ on $z$ is going to determine the effective layer thicknesses  and average volume fractions for each layer used in order to approximate the continuous variation of $f(z)$ as depicted schematically in Fig.~\ref{SchemeLayers}. Each layer has its  corresponding $\epsilon^{eff}(\omega)$ obtained by solving eq. (\ref{Bruggeman}). This is the second approximation we introduce into the model, the continuously varying effective dielectric function $\eps(\omega,z)$ is approximated by a $z$-independent function within each layer. This discretization is a well known procedure to solve electromagnetic problems involving continuously variable  inhomogeneous media for which there is no analytical solution~\cite{Chew95}.

Therefore, in our model we reduce the task of calculating the Casimir force between two rough surfaces separated by the average gap $d$ to that of calculating the force between multilayered systems separated by an effective gap $l$ (see fig.~\ref{SchemeLayers}). Currently, this calculation can only be performed for the case of planar geometry using the expressions for the pressure produced by the Casimir effect derived by Toma\u{s}~\cite{Tomas02} and Raabe {\it et al.}~\cite{Raabe03} as generalizations of the Lifshitz theory~\cite{Lifshitz56}, which in the more general case of nonzero temperature $T$ can be cast in the form~\cite{Raabe03}
\be
P(l) = \frac{k_{\mathrm{B}} T}{\pi} {\sum_{m=0}^\infty}^\prime
 \int_0^\infty{dq q \kappa \sum_{\sigma=s,p}}  \frac{r_-^\sigma r_+^\sigma e^{-2 \kappa l}}{1-r_-^\sigma r_+^\sigma e^{-2 \kappa l}}\,,
\label{Pc}
\ee
where the term for $m=0$ must be multiplied by $1/2$. In this equation $\kappa = \sqrt{\xi^2_m \eps(i \xi_m)/c^2+ q^2}$, with the Matsubara frequencies $\xi_m=2 m \pi k_{\mathrm{B}} T/\hbar$. $r_\pm^\sigma=r_\pm^\sigma(q,\xi)$ are the generalized Fresnel reflection coefficients for $s$ and $p$ polarized waves reflecting from the stack of layers above (subscript $+$) or below (subscript $-$) the effective gap $l$. In the case of a vacuum gap $\eps(i \xi) = 1$, however, it is worth noting that, in general, $\kappa$ is a function of the dielectric function calculated over the imaginary frequency axes. The Fresnel reflection coefficients can be easily obtained, for instance, from the set of recurrence relations derived in ref. \cite{Raabe03}, and result to be functions of the layers thicknesses $d_{n\pm}$ and their effective dielectric functions $\eps_{n\pm}(i \xi)$.  

Before we present the results for the Casimir force predicted by the MEMM it is worth discussing its expected range of applicability. The relevant parameters for the analysis are  $\sigma = \sqrt{\left\langle h(x,y)^2 \right\rangle}$, $\Lambda$, the average surface gradient $\left\langle \mid \nabla  h(x,y) \mid \right\rangle$ at each surface, and the mean gap $d$. For the sake of brevity in the discussion we assume all parameters to be at least approximately the same for both interacting surfaces, but the discussion could focus, for instance, on the surface with the largest $\sigma$ or $\Lambda$. We start by noting that the EMA is considered to correctly represent the properties of a dielectric medium with inclusions in a host material whose dimensions are small compared to the wavelength $\lambda$ of the incident electromagnetic wave. It is assumed conservatively that the largest dimension (height, radius or sides of the surface features), $l_{max}$, must satisfy $l_{max} \lesssim \lambda/10$~\cite{Sihvola00,Sihvola08}.  Due to the symmetric treatment of inclusions and host material the Bruggeman model has the ability to model the electrical response of the clusters and more complex aggregates actually found at a rough surface. For spherical inclusions such larger structures begin to form close to the percolation threshold of $f \sim 0.3$, when the randomly distributed spheres get into contact with neighbouring spheres forming a geometrically connected phase~\cite{Garboczi95}. When such larger structures are present it is the average size of the surface structures that becomes relevant. Considering the symmetric treatment of voids and solid the average lateral and vertical dimensions of such structures are of the order of $\Lambda$ and  $\sigma$, respectively. Now, we have to consider that the vacuum electromagnetic modes relevant to the Casimir effect are those with a wavelength $\lambda$ satisfying $\lambda \sim d$. Therefore, the following inequality must be satisfied $l_{max} \lesssim 0.1 \lambda \approx  0.1 d$, where $l_{max} = Max[\sigma,\Lambda]$. This is a conservative limit on $l_{max}$ compared to other limits found in the literature~\cite{Sirvent11} and the actual limit could be less stringent. For this reason the above inequality can still be considered as an adequate criteria  when the roughness of both surfaces is relevant. However, further theoretical investigations should be performed to set more precisely the range of validity of the model. To conclude, it is worth noting that differing from the other methods of calculation, for the proposed model there are no upper limits imposed on $\mid \nabla h(x,y) \mid$. As a consequence, the  restriction that $\sigma \ll \Lambda$ does not apply, demonstrating that the MEMM can be used for short scale roughness. 

For the sake of concreteness let us consider the implications of the expected range of validity of the model  in the  context of small gap micro- and nanodevices, where the precise knowledge of the roughness effect on  the Casimir force can be more relevant. For such devices it is generally the case that the Casimir force becomes relevant when $d \lesssim 100$ nm~\cite{Gusso,Gusso07,Pruvost09}. In this case, the restriction $l_{max} \lesssim 0.1 d$ implies that $\Lambda$ and $\sigma$ at the rough surface must be smaller than approximately 10 nm. Considering that $\sigma$ is expected to be of the order of a few nanometers results that $\left\langle \mid \nabla  h(x,y) \mid \right\rangle$ is of the order unity. While well suited for the use of the MEMM, this condition can be considered out of the range of applicability of the PWS, PFA and PSA.

While no further considerations should be made in the case of dielectric materials, for metals the situation is more involved. There are two other restrictions. One results from the fact that for the validity of the EMA the penetration depth $\delta$ of the electromagnetic waves must be large compared to the average dimension of the surface features. This restriction results from the exponential decay of the electromagnetic fields inside lossy inclusions such as metals~\cite{Sihvola00}. As a consequence, the following inequality must be satisfied $l_{max} \lesssim \delta$, where $\delta \sim 20$ nm for good conductors. This is essentially the same restriction expected in the context of small gap micro and nanodevices derived above.  The other restriction is related to the potential effects of spatial dispersion on the determination of the Casimir force~\cite{Bordag09}. The effect of spatial dispersion in metals can be neglected if the condition $v_F/\omega = v_F \lambda/(2 \pi c) \ll l_{max}$ is satisfied~\cite{Klimchitskaya00}, where $v_F$ is the electron Fermi velocity. This condition assures that the electrons are going to oscillate restricted to a small region of the effective layer were the dielectric function does not vary appreciably.  As a reference, for gold $v_F = 1.4\times 10^6$ ms$^{-1}$, and this restriction implies the approximate inequality $10^{-3} d \ll l_{max}$. This restriction is less stringent than the previous ones. 

\begin{figure}
\centerline{\includegraphics[scale=0.45]{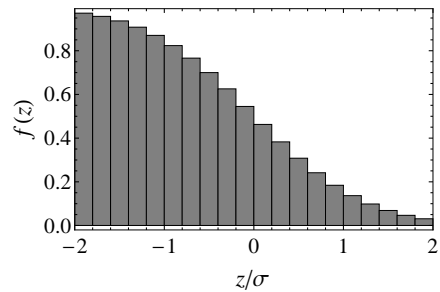}}
\caption{Histogram of the function $f(z)$ for a Gaussian height distribution.}
\label{histo}
\end{figure}

\section{Results}

In what follows we are going to present illustrative results for the Casimir force predicted by the MEMM. We consider the planar configuration where two rough surfaces described by the functions $h_{-}(x,y)$ and $h_{+}(x,y)$, have an average separation gap $d$ (See fig.~\ref{SchemeLayers}). In the MEMM the only required information regarding the rough surface are $\sigma$ and the amplitude probability density function $p(h)$.   We consider the two surfaces having a Gaussian height distribution
\be
p(h) = \frac{1}{\sqrt{2 \pi} \sigma} e^{-\frac{h^2}{2\sigma^2}} \,,
\ee
with the same rms amplitude $\sigma$.  From $p(h)$ the volume fraction (material ratio) function $f(z)$ can be calculated from the cumulative distribution function as~\cite{Whitehouse03}
\be 
f(z) = 1-\int_{-\infty}^z p(z')dz'= \frac{1}{2}\left[1-Erf\left( \frac{z}{\sqrt{2}\sigma} \right)\right]\,. 
\ee
This function varies monotonically from $f=1$ (bulk) down to $f=0$ (vacuum) as $z$ goes from large negative up to large positive values. Most of the variation occurs within the interval $|z|< 2 \sigma$ as can be inferred from fig.~\ref{histo}, where the average values of $f(z)$ within intervals of $\Delta z = 0.2 \sigma$ are plotted. 

\begin{table}
\caption{Rough layer models. Presented are the number of sublayers, layer thickness and volume fraction.}
\label{table1} 
\begin{center}
\begin{tabular}{c c c c }
Model &  $\sharp$ of Sublayers  &  Layer thickness & $f$  \\
1 & 1 & $1.2\sigma$ &  0.5 \\
2 & 1 & $1.6\sigma$ & 0.5 \\
3 & 2 & $0.8\sigma/0.8\sigma$ & 0.35/0.65 \\
4 & 3 & $0.4\sigma/1.2\sigma/0.4\sigma$ & 0.2/0.5/0.8 \\
\end{tabular}
\end{center}
\end{table}

For the Gaussian surfaces we are considering, with the increase in the number of layers, a good convergence was already obtained when the rough layer was modeled by three layers, resulting in a system comprised of nine layers. We have tested several models, comprised of one or more layers laying between the vacuum gap and a semispace having the optical properties of the bulk solid, and choose to present the predictions based on four models. The models are presented in table~\ref{table1}, and are applied for both surfaces $h_{-}(x,y)$ and $h_{+}(x,y)$, resulting in a symmetric system with $d_{n+}=d_{n-}$, as well as, $f_{n+}=f_{n-}$. 

\begin{figure}
\centerline{\includegraphics[scale=0.5]{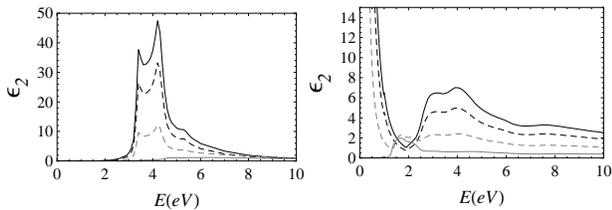}}
\caption{$\eps_2$ as a function of photon energy $E$ for the bulk (black/full), and effective medium with $f=0.8$ (black/dashed), $f=0.5$ (gray/dashed), and $f=0.2$ (gray/full). The left and right panels are for Si and Au, respectively.}
\label{eps2}
\end{figure}

In fig.~\ref{eps2} we present illustrative results for  the imaginary part of $\eps^{eff}(\omega)$ predicted by the Bruggeman model and used to calculate $\eps^{eff}(i \xi)$ for each layer. The results are for silicon (Si) and gold (Au), and were obtained using the experimental data on both $\eps_1={\rm Re}[\eps(\omega)]$ and $\eps_2={\rm Im}[\eps(\omega)]$ as given by Adachi~\cite{Adachi99} and Palik~\cite{Palik85}, respectively. It can be seen that $\eps_2(\omega)$ for silicon varies smoothly as $f$ increases from values close to zero, characterizing the prevalence of the contribution from the vacuum, over those of the bulk. In the case of gold a more complex result is evidenced. Below the percolation threshold of $f = 1/3$, $\eps_2(\omega)$ is essentially that expected for an insulator (see, for instance, the curve for $f=0.2$). Above the percolation threshold a metallic behavior is observed. It is characterized by the divergence of $\eps_2(\omega)$ as $\omega$ tends to zero. It is also worth  noting that for $f \gtrsim 0.8$ and $f \lesssim 0.2$ we have the approximate results $\eps^{eff}_2(\omega) \approx \eps^{bulk}_2(\omega)$ and $\eps^{eff}_2(\omega) \approx 0$, for both metals and semiconductors. This fact can be used to determine the most adequate thickness to be considered for the effective medium region since   the effect on the Casimir force resulting from the regions where $f \gtrsim 0.8$ and $f \lesssim 0.2$ can be approximated by those of the bulk and vacuum, respectively. That is the reason why in the models 2, 3 and 4 we have considered a roughness layer with thickness ranging from $1.6 \sigma$ up to $2 \sigma$,  encompassing the region that most affects the Casimir force. Model 1 was considered for the sake of comparison.

In fig.~\ref{Figeta}, we present the roughness correction factor $\eta_r(d)$ to the Casimir force between two semispaces made from Si and Au. This correction factor singles out the roughness effect from the total force that includes the temperature and finite conductivity corrections. The results are for a temperature $T= 300$ K and a value of $\sigma = 5$ nm was chosen in order to illustrate the effect of large amplitude roughness on the Casimir force. We investigated the predicted Casimir pressure for Au  considering $\eps(\omega)$ described by both the plasma and Drude models. The results presented in fig.~\ref{Figeta} are those obtained using the plasma model. $\eta_r(d)$ calculated using the Drude model at $T = 0$ K differs only slightly (by less than approximately $2\%$) from these results. Due to the potential effects of spatial dispersion at large separations, discussed previously, the calculation of $\eta_r(d)$ for Au was restricted to shorter separations. 

In the calculations based on the Drude model we extended the experimental data~\cite{Palik85} on $\eps(\omega)$ to lower frequencies adopting $\omega_p = 9.0$ eV and $\gamma = 0.035$ eV for the plasma and relaxation frequencies. Using eq. (\ref{Bruggeman}) the effective $\eps_2^{eff}(\omega)$ is calculated directly from $\eps^{D}_1(\omega)$ and $\eps^{D}_2(\omega)$ predicted by the Drude model for the bulk and introduced into the usual Kramers-Kronig relation giving $\eps^{eff}(i \xi)$ for the effective layers. For the plasma model, the calculations are more involved.  In this case $\eps^{p}_2(\omega)$ predicted by the plasma model for the bulk is obtained by subtracting from $\eps^{D}_2(\omega)$ the contribution from the relaxation of the conduction electrons. The available experimental data for crystalline gold presents inconsistencies because they result from the combination of different experimental data sets~\cite{Palik85}. For this reason, in order to generate a physically consistent $\eps^{p}(\omega)$, $\eps^{p}_1(\omega)$ was obtained from $\eps^{p}_2(\omega)$ using the generalized Kramers-Kronig relation~\cite{Klimchitskaya07}
\be
\eps_1(\omega) = 1 + \frac{2}{\pi}P\int_0^{\infty} \frac{\xi \eps_2(\xi)}{\xi^2-\omega^2}d\xi -\frac{\omega_p^2}{\omega^2} \,,
\ee
with $\omega_p$ given above. Finally, $\eps_2^{eff}(\omega)$ is determined from eq. (\ref{Bruggeman}) and used to calculate $\eps^{eff}(i \xi)$ by means of the generalized Kramers-Kronig relation whenever $f>1/3$. This procedure is necessary because of the resulting metallic behavior of the effective medium with the associated  plasma frequency  given by $(\omega_p^{eff})^2=(3f-1)\omega_p^2/2$. For a volume fraction below the percolation threshold the effective medium has an insulatorlike dielectric function, and the usual Kramers-Kronig relation can be used in order to calculate $\eps^{eff}(i \xi)$.

\section{Comparison with PFA and PSA}

\begin{figure}
\centerline{\includegraphics[scale=0.56]{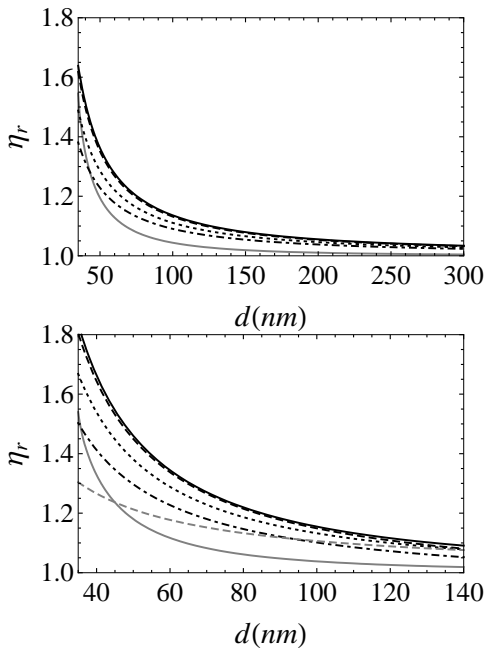}}
\caption{$\eta_r(d)$ for the models in table~\ref{table1}: model 1 (black/dot-dashed), 2 (black/dashed), 3  (black/dotted), and 4 (black/full). Also presented are the predictions based on the PFA (gray/full) and PSA (gray/dashed). The left and rigth panels are for Si and Au, respectively.}
\label{Figeta}
\end{figure}

For the sake of comparison $\eta_r(d)$ predicted based on the PFA and PSA are also presented in fig.~\ref{Figeta}. 
While approximate analytical expressions for $\eta_r^{\rm PFA}(d)$ can be derived, we resorted to the numerical calculation of  $\eta_r^{\rm PFA}(d)$. A set of synthetic Gaussian surfaces was generated and $\eta_r^{\rm PFA}(d)$ calculated for an ensemble of surface pairs according to the following expression
\be
\eta_r^{\rm PFA}(d) = \frac{d^4}{\eta_{P}(d) A} \iint_A dA \frac{\eta_{P}(d+\delta h)}{(d+\delta h)^4}\,,
\label{pfa} 
\ee
where $\delta h = \delta h(x,y) = h_{+}(x,y)-h_{-}(x,y)$, $A$ is the surface area, and $\eta_{P}(d)$ is the finite conductivity and temperature correction factor for the pressure between two semispaces having plane surfaces. The use of synthetic surfaces allowed us to establish a more realistic scenario setting, for instance, the distance of contact between surfaces, which for the  ensemble of surface pairs with $\sigma = 5$ nm was $d \sim 33$ nm. For this reason, our results are restricted to $d > 35$ nm. We note that for stochastic rough surfaces the Casimir force and, consequently, 
$\eta_r$ predicted by the PFA are only functions of the amplitude probability density of the interacting rough surfaces (see Section 17.2.2 of ref.~\cite{Bordag09}) as for the MEMM. However, while these two methods have no explicit dependence on other roughness parameters, both $\Lambda$ and $\langle \mid \nabla h(x,y) \mid \rangle$, for instance, must be known and taken into account to determine which method could provide the most reliable predictions.     
From the results presented in fig.~\ref{Figeta} we conclude that the effects of short scale roughness are underestimated by the PFA when compared to the MEMM. Such discrepancy should be expected due to the distinct ranges of validity of these approaches. While in both approaches the surfaces are described solely by the same amplitude probability density, the actual roughness profiles that could be accounted by each model are quite different. For the MEMM the surface must have a much  smaller $\Lambda$ (more specifically $\Lambda \ll d$) corresponding to a much more compact roughness profile, which is seen by the relevant electromagnetic waves as a flat layer composed of an effective material whose physical properties are between those of the vacuum and the bulk. For the validity of PFA, the rough surface must be considered as piecewise plane, each piece being described as two interacting semispaces, a physical picture that differs quite significantly from that considered for the MEMM. 

For a comparison with the PSA we resorted to approximate analytical results. The roughness correction factor $\eta_r^{PSA}$ was calculated based on the approximate analytical result for the roughness correction predicted for a metal described by the plasma model, namely
\be
\eta^{PSA}_r(d) = 1+\frac{4}{3}\Delta\,.
\label{PPSA}  
\ee
In this equation the  Casimir energy relative correction factor $\delta E/E =\Delta = 2.7 \sqrt{\pi} \sigma^2/(\Lambda d)$ was derived in ref.~\cite{MaiaNeto05} under the condition $\Lambda \ll d \ll \lambda_p$, where $\lambda_p = 136$ nm is the plasma wavelength of gold. The same condition is valid for eq. (\ref{PPSA}), which in the case of a surface covered with short scale roughness  can only be approximately satisfied if the condition $\sigma \ll \lambda$ is to be kept. Only as an illustrative result, we push the predictions based on the PSA slightly beyond  the limits of its range of applicability and keeping $\sigma =5$ nm we assume a short scale roughness with $\Lambda = 3 \sigma = 15$ nm in order to plot the curve presented  in fig.~\ref{Figeta}. Considering the range of validity of eq. (\ref{PPSA}) the result is approximately valid in the region around $d = 60$ nm. For the sake of comparison, while the PSA  predicts $\eta^r(d) \propto d^{-1}$, fitting $\eta^r(d)$ for model 4 leads to the conclusion that $\eta^r(d) \propto d^{-1.6}$ for Au. The exponent $\alpha = 1.6$ is between those predicted by the PSA ($1\leq \alpha \leq 2$) for different ranges of the relevant parameters~\cite{MaiaNeto05}. Furthermore, reproducing the trends observed in ref. \cite{MaiaNeto05},  eq.(\ref{PPSA}) predicts a roughness correction larger than that of PFA in a wide range of separations. In fact, the approximate result eq. (\ref{PPSA}) clearly demonstrates that the smaller the $\Lambda$ the larger the roughness correction, a trend also evidenced in ref.~\cite{MaiaNeto05} by numerical calculations and further approximate analytical results.  However, for the chosen values of $\sigma$ and $\Lambda$ the condition $\langle \mid h(x,y) \mid \rangle \ll 1$ can not be adequately satisfied and we should rely only on the prediction  based on the MEMM. Following the trend indicated by the approximate PSA results  the MEMM predicts an even larger roughness correction. This result evidences the actual relevance of short scale roughness to the Casimir force.   

\begin{figure}
\centerline{\includegraphics[scale=0.45]{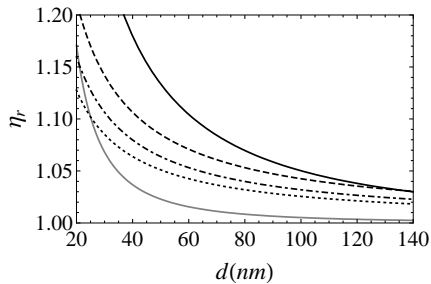}}
\caption{$\eta_r(d)$ for a Gaussian surface roughness with $\sigma = 2$ nm. The curves are the predictions based on the model 4 for the MEMM (black/full), PSA with $\Lambda = 3 \sigma$ (dashed), $\Lambda = 4 \sigma$ (dot-dashed), and $\Lambda = 5 \sigma$  (dotted), and PFA (gray/full).}
\label{Figeta2nm}
\end{figure}

Due to the importance of correctly establishing the range of applicability  of the MEMM we further compare its predictions with those of the PSA. In order to obtain reliable predictions from the approximate result eq. (\ref{PPSA}) over a wider range of the separation $d$ we consider a surface with a smaller roughness amplitude. In fig.~\ref{Figeta2nm}  we present the predictions from the PFA, PSA and the MEMM for gold surfaces having $\sigma = 2$ nm. Therefore, the condition $\Lambda \ll d \ll \lambda_p$ can be more appropriatelly satisfied over a wider range of separations $d$, simultaneously with the constraint $\sigma \ll \Lambda$, also required for the validity of the PSA. In fig.~\ref{Figeta2nm} we present $\eta_r^{PSA}(d)$  calculated for $\Lambda = 3 \sigma, 4 \sigma$ and $5 \sigma$. For such values of $\sigma$ and $\Lambda$ the approximate results for the PSA are valid in the range $20 \lesssim d \lesssim 60$ nm.  The approximate results of the PSA indicate a qualitative convergence towards the prediction of the MEMM before the condition $\sigma \ll \Lambda$ ceases completely to be valid.  Therefore, the comparison between the predictions of PSA and that of the MEMM indicates that the proposed model can be used whenever $\Lambda \lesssim 3 \sigma$, well within the range of definition of short scale roughness.  We can expect that the comparison between the predictions based on improved models, such as higher order PSA, and the MEMM will give further information regarding the range of validity of the proposed model. However, we can further advance that the MEMM based on the Bruggeman mixing rule, eq. (\ref{Bruggeman}), can not be expected to give reliable predictions for rough surfaces characterized by having $\Lambda \ll \sigma$ and $\Lambda \gg \sigma$. In such limits the structures found at the rough surface are seen by the electromagnetic waves as essentially  one and two-dimensional structures, respectively, and the Bruggeman mixing rule is limited to describe the effects of fully three-dimensional structures. It can be assumed, conservatively, that the predictions of the MEMM based on the Bruggeman mixing rule are valid within the range $\sigma/3 \lesssim \Lambda \lesssim 3 \sigma$.  
\section{Conclusion}
To conclude, it is worth to note that the predictions based on the single layer model (model 2) are approximately the same as those predicted by the more complex three layer model (model 4). This conclusion, being the same for Si and Au, suggests that even a single layer model can accurately represent the effective properties of  rough layers. Therefore,  with a relatively small calculational effort, the MEMM can deliver accurate predictions of the Casimir force when short scale roughness is involved. Furthermore, by comparing the results for other three and four layer models we observed a small variation on the predicted force, within $\pm 5 \%$, in the expected range of validity of the model. Finally, while the large roughness correction predicted by the MEMM  is in qualitative agreement with the experimental results of ref.~\cite{Zwol10}, further theoretical and experimental investigations are required in order to stablish the range of validity of the model.
\begin{acknowledgments}
A. G. is thankful to P. A. Maia Neto for helpful discussions and suggestions. This work was supported by the  Conselho Nacional de Desenvolvimento Cient\'ifico e Tecnol\'ogico, CNPq-Brazil, and Funda\c{c}\~ao Carlos Chagas Filho de Amparo \`a Pesquisa do Estado do Rio de Janeiro-FAPERJ. 
\end{acknowledgments}

\end{document}